\documentclass[a4paper,prl,twocolumn,floatfix,footinbib]{revtex4}
\usepackage{amsmath,amssymb,mathrsfs}
\usepackage{graphicx}

\begin{document}
\newcommand{\ds}{\displaystyle}
\newcommand{\ud}{{\mathrm{d}}}
\newcommand{\us}{{\mathrm{s}}}
\newcommand{\ubi}{{\mathrm{bi}}}
\newcommand{\LD}{L_{\mathrm{D}}}
\newcommand{\uc}{{\mathrm{c}}}
\newcommand{\ub}{{\mathrm{b}}}
\newcommand{\uB}{{\mathrm{B}}}
\newcommand{\uf}{{\mathrm{f}}}

\title{Scaling of nano-Schottky-diodes}

\author{G.D.J. Smit}
\author{S. Rogge}
\author{T.M. Klapwijk}
\affiliation{Department of Applied Physics and DIMES, Delft
University of Technology, Lorentzweg 1, 2628 CJ Delft, The
Netherlands}

\begin{abstract}
A generally applicable model is presented to describe the
potential barrier shape in ultra small Schottky diodes. It is
shown that for diodes smaller than a characteristic length $l_\uc$
(associated with the semiconductor doping level) the conventional
description no longer holds. For such small diodes the Schottky
barrier thickness decreases with decreasing diode size. As a
consequence, the resistance of the diode is strongly reduced, due
to enhanced tunneling. Without the necessity of assuming a reduced
(non-bulk) Schottky barrier height, this effect provides an
explanation for several experimental observations of enhanced
conduction in small Schottky diodes.
\end{abstract}

\date{\today}

\maketitle

The effect of downscaling the dimensions of a device on its
electrical transport properties is an important topic today.
Extremely small diodes have been experimentally realized and
characterized in various systems, e.g. carbon nanotube
heterojunctions \cite{yao99}, junctions between $p$-type and
$n$-type Si nanowires \cite{cui01} or junctions between the
metallic tip of a scanning tunneling microscope and a
semiconductor surface \cite{avouris93,hasunuma98}.  These
experiments showed several deviations from conventional diode
behavior. Despite some modelling in truly one-dimensional systems
\cite{leonard99,odintsov00}, little work has been done on
modelling the effects of downscaling a conventional diode, in the
regime where quantum confinement does not play a role.

In this paper we present a simple model (based on the Poisson
equation) describing the barrier shape in a diode, that is readily
applicable to arbitrarily shaped small junctions. It is related to
descriptions of inhomogeneities in the Schottky barrier height
(SBH) in large diodes \cite{tung92}, barrier shapes in small
semiconducting grains \cite{malagu02} and charge transfer to
supported metal particles \cite{ioannides96}. Although we restrict
ourselves to metal-semiconductor junctions, the model can easily
be adapted for e.g. $p$-$n$-junctions. The main result is that if
the size of the metal-semiconductor interface is smaller than a
characteristic length $l_\uc$, the thickness of the barrier is no
longer determined by the doping level or the free carrier
concentration, but instead by the size and shape of the diode. The
resulting thin barrier in small diodes will give rise to enhanced
tunneling, qualitatively explaining measurements of enhanced
conduction \cite{avouris93,hasunuma98,smit02}, without the
necessity of assuming a reduced SBH. Moreover, experimentally
observed scaling behavior and deviating $IV$-curve shapes
\cite{smit02} can be explained.

The transport properties of a Schottky diode are governed by the
potential landscape which has to be traversed by the charge
carriers. First, we study an easily scalable and highly
symmetrical model system, namely a metallic sphere embedded in
semiconductor (see Figure~1, upper left inset). The radius $a$ of
the metallic sphere is a measure for the interface size: for large
$a$, we expect to find the well known results for a conventional
diode, while decreasing $a$ gives the opportunity to study finite
size effects.

We only model the barrier shape in the semiconductor; the SBH
$\varphi_\uB$ is accounted for in boundary conditions and is
considered as a given quantity. For simplicity, the depletion
approximation \cite{sze81} is adopted, which is valid for a wide
range of realistic parameters. Moreover, the space charge region
is assumed to be homogeneously charged, an assumption that will be
discussed later. Solving the Poisson equation in $n$-type silicon
with the boundary condition that the charge on the sphere cancels
the total charge in the space charge region, we find for $0\leq
x\leq w$
\begin{equation}
  \begin{array}{l}
    \ds\frac{e}{kT}\cdot V(x)=\\
    \\
    \ \ \ \ \ds\frac{1}{2\LD^2}\left[(a+w)^2-
    \ds\frac{2(a+w)^3}{3(a+x)}-\frac{(a+x)^2}{3}\right],\\
  \end{array}
  \label{eq:vxnorm}
\end{equation}
where $x$ is the radial distance from the interface, $w$ the
depletion width and $\LD=\sqrt{\varepsilon_\us kT/(e^2N_\ud)}$ the
Debye length. The zero-point of the potential is chosen in the
semiconductor bulk. The value of $w$ is fixed by the second
boundary condition $V(0)=V_\us$, where $V_\us$ is the total
potential drop over the space charge region and satisfies
$V_\us=(\varphi_\uB-\varphi_\us)/e-V$ (with
$\varphi_\us=E_\uc-E_\uf$). Eq.~(\ref{eq:vxnorm}) is valid for
small bias voltage $V$. The limited validity of the depletion
approximation at finite temperatures only affects the tail of the
barrier (where $|V(x)|\lesssim kT$), which is unimportant for the
transport properties. From the equation, it can be seen that the
characteristic length scale of this system is
$$
  l_\uc\stackrel{\mathrm{def}}{=}
  \LD\sqrt{2eV_\us/kT}=\sqrt{\frac{2\varepsilon_\us V_\us}{eN_\ud}}.
$$
By comparing the diode size $a$ to $l_\uc$ we can decide whether
the diode is `small' or `large'. In the lower right inset of
Figure~1 the value of $l_\uc$ is plotted versus doping
concentration $N_\ud$.

An important quantity for electrical transport is the Schottky
barrier thickness. In Figure~1, the barrier full width at half
maximum (FWHM, $x_{1/2}$) calculated from Eq.~(\ref{eq:vxnorm}) is
plotted as a function of diode size $a$. From the figure it is
clear that for $a\gg l_\uc$ the value of $x_{1/2}$ approaches a
constant, which was expected for a large diode. Indeed, for $a\gg
l_\uc$, Eq.~(\ref{eq:vxnorm}) reduces to
$V(x)=-\frac{eN_\ud}{2\varepsilon_\us}(x-w)^2$, which is the
well-known textbook \cite{sze81} result for band bending in the
depletion approximation for an infinitely large diode. Both the
depletion width $w=\sqrt{(2\varepsilon_\us/eN_\ud)V_\us}$ and
$x_{1/2}$ are in that regime independent of $a$.

\begin{figure}
  \centering
  \includegraphics[width=8cm]{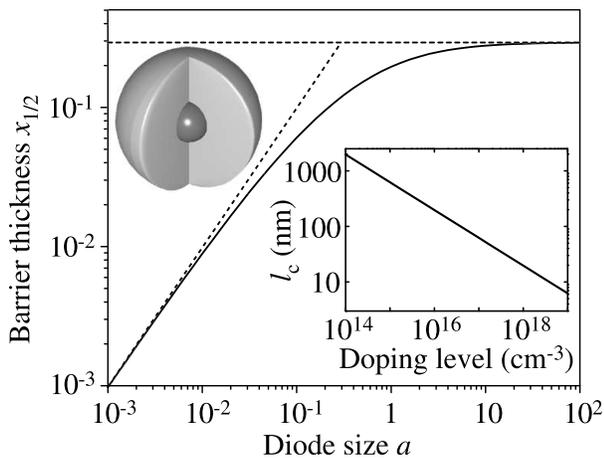}
  \caption{Plot of the calculated barrier FWHM $x_{1/2}$
  as a function of diode size $a$ (based on Eq.~(\ref{eq:vxnorm})),
  both in units of $l_\uc$. The dashed lines represent the
  asymptotic values for $a\gg l_\uc$ (conventional diode) and $a\ll
  l_\uc$ (new regime) respectively. The lower right inset is a plot
  of $l_\uc$ as a function of doping level $N_\ud$ in silicon
  ($\varepsilon_\us=11.7$) for $\varphi_\uB=0.67$~eV and $T=300$~K.
  The upper left inset schematically shows the model system, a
  metallic sphere embedded in semiconductor.}
\end{figure}

Figure~1 shows that for $a\lesssim l_\uc$ the value of $x_{1/2}$
is no longer constant, but decreases with decreasing $a$. For
$a\ll l_\uc$ it approaches $x_{1/2}=a$, i.e.~the barrier thickness
equals the diode size. This also follows from
Eq.~(\ref{eq:vxnorm}), which reduces to $V(x)=V_\us\cdot a/(a+x)$
for $a\ll l_\uc$ and $x\ll w$ (that is, close to the interface).
Note that this is exactly the potential due to the charged sphere
only. In this regime, the effect of the semiconductor space charge
on the barrier shape and thickness can be neglected. This can be
understood from the fact that the screening due to the space
charge region takes place on a length scale $l_\uc$, as in
conventional (large) diodes. However, from Gauss's law it follows
that any charged object of typical size $d<\infty$ in a dielectric
medium gives rise to a potential that behaves roughly as
$V(r)\propto d/r$. This Coulomb potential can be further screened
by the formation of a space charge layer of opposite sign, but
that additional screening can be neglected if $d\ll l_\uc$. This
observation holds for \emph{any} interface with typical dimensions
much smaller than $l_\uc$.

\begin{figure}
  \centering
  \includegraphics[width=8cm]{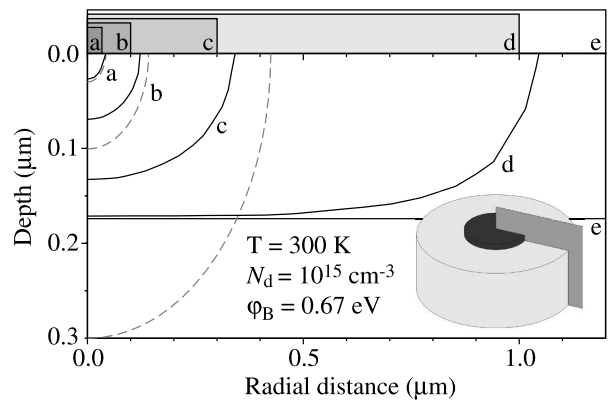}
  \caption{The solid lines are contours of the barrier
  FWHM for various disc-shaped contacts (see inset; radii ranging
  from 30\,nm (a) to infinite (e)), taken from a numerical solution
  of the Poisson equation in silicon. It clearly shows the contact
  size dependence for contact radii smaller than $l_\uc \sim
  750$~nm. The dashed lines are the FWHM-contours of the barrier for
  the three smallest diodes, neglecting the screening effect of the
  semiconductor space charge region. The inset indicates the plane
  of cross-section shown in the figure.}
\end{figure}

In a geometry that can actually be fabricated, the Poisson
equation must be solved numerically. We have done this for
$n$-doped silicon ($N_\ud=10^{15}\,\mathrm{cm}^{-3}$) in contact
with metallic circular disks of various radii. In all further
calculations $\varphi_\uB=0.67\,$eV was used, which is the barrier
height of the CoSi$_2$/Si(111)-interface \cite{tung92b}. Figure~2
shows the FWHM-contours of the barriers as resulting from these
calculations. Also shown are the FWHM-contours of the barrier due
to the metallic contacts only, illustrating the negligible effect
of the space charge region on the barrier thickness in very small
diodes \cite{footnote1}.

To study the effect of the reduced barrier width on the transport
properties of a small Schottky diode, a transmission coefficient
$T(E,V)$ was obtained for the barrier shape from
Eq.~(\ref{eq:vxnorm}). This was done in a one-dimensional fully
quantum mechanical calculation \cite{walker94}. Note that $T(E,V)$
is implicitly dependent on temperature and doping level, because
these quantities influence the position of the Fermi-level in the
bulk semiconductor. The current density is then given by
$$
  J(V)\propto\int_0^\infty T(E,V)[f(\varphi_s+E)-f(\varphi_s+E+V)]
        \,\mathrm{d}E,
$$
from which it follows that the zero bias differential conductance
satisfies
$$
  \left.\frac{\ud J}{\ud V}\right|_{V=0}\propto
  -\int_0^\infty T(E,V)f'(\varphi_s+E)\,\mathrm{d}E.
$$
Here, $f$ is the Fermi-Dirac distribution function and $E$ the
energy above the semiconductor conduction band edge. Transport due
to electrons at energies below the barrier maximum ($E<V_\us$) is
regarded as tunneling, while for $E>V_\us$ we speak of thermionic
emission. Obviously, the contribution of thermionic emission is
almost independent of the barrier thickness, while tunneling is
strongly dependent on the barrier thickness.

\begin{figure}
  \centering
  \includegraphics[width=8cm]{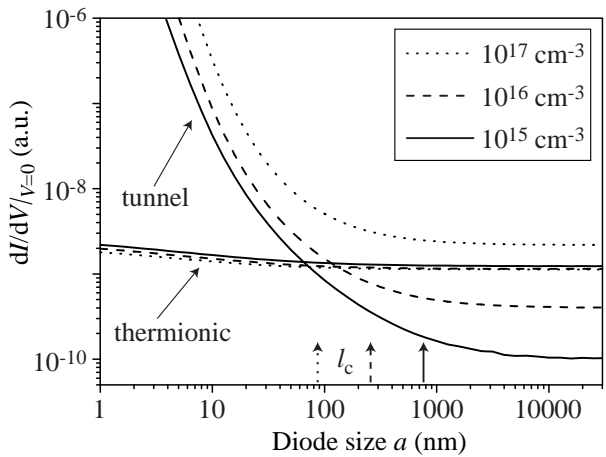}
  \caption{The contributions of tunneling and thermionic
  emission to the zero bias differential conductance, plotted as a
  function of diode size $a$ for various doping concentrations. The
  vertical arrows indicate the values of $l_\uc$. The parameters are
  the same as in Figure~2.}
\end{figure}

In Figure~3, the calculated zero bias differential conduction is
plotted as a function of diode size $a$ for several values of
$N_\ud$. For $a\gtrsim l_\uc$ this quantity is independent of $a$.
For smaller values, the tunnel current starts to increase rapidly,
eventually leading to a strong increase of the total conduction.

\begin{figure}
  \centering
  \includegraphics[width=8cm]{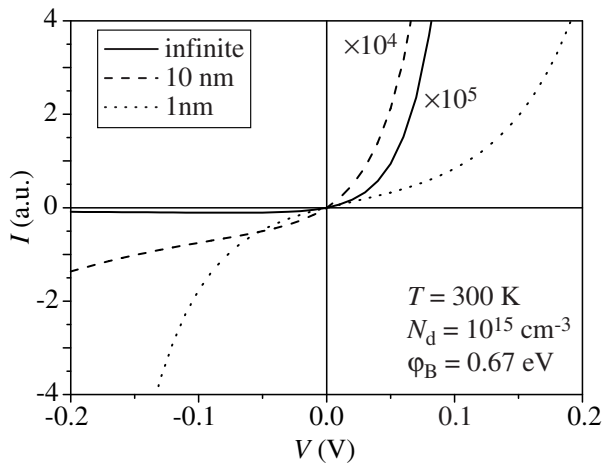}
  \caption{Calculated $IV$-curves for various diode
  sizes. The large diode curve has the expected exponential shape.
  The qualitative appearance of the curves changes drastically with
  decreasing diode size. The curves of the larger diodes have been
  scaled vertically.}
\end{figure}

Moreover, the shape of the $IV$-curves changes with decreasing
diode size. Our calculations (Fig.~4) show that for large diodes
the $IV$-curve has exactly its expected exponential shape
($I\propto [\exp(eV/kT)-1]$). Apart from the total current
increase, in small diodes the relative contribution of the reverse
current starts to increase and eventually---in extremely small
diodes---the reverse current exceeds the forward current, thus
reversing the rectifying behavior of the diode.

Note that the image charge effect \cite{sze81} was neglected so
far. However, inclusion of this effect would only enhance the
phenomenon mentioned above, as it reduces the effective barrier
height and width even further, especially in barriers which are
narrow already.

One more issue that needs to be discussed is that of discrete
random dopants. In our analysis, the dopants played a role in
determining the Fermi-level position in the semiconductor bulk and
were considered to provide a homogeneous space charge region.
However, for the realistic parameters $N_\ud=10^{15}\,$cm$^{-3}$
and $a=30\,$nm (so $a\ll l_\uc$) the volume in which the potential
drops to half its initial value contains approximately one doping
atom. Discrete energy levels of such a doping atom cannot be
resolved at room temperature. More importantly, the potential well
due to an ionized single dopant will locally distort the barrier
shape. This effect complicates the potential landscape, but it can
only significantly increase the conduction of the diode, when the
dopant resides close to the interface \cite{footnote2}.

In conclusion, we have shown by means of a simple electrostatic
argument that the Schottky barrier thickness becomes a function of
the diode size for small diodes (e.g. smaller than $l_\uc\approx
80\,$nm for $N_\ud=10^{17}\,$cm$^{-3}$). Consequently, the
contribution of tunneling to the total conductance is greatly
enhanced in small diodes. This effect explains several
experimental results \cite{avouris93,hasunuma98}, without the
assumption of a reduced SBH. Moreover, small diodes show
$IV$-curve shapes that qualitatively differ from those of
conventional diodes.

We wish to thank J.~Caro and H.W.M.~Salemink for detailed
discussions concerning this work. This work is part of the
research programme of the 'Stichting voor Fundamenteel Onderzoek
der Materie (FOM)', which is financially supported by the
'Nederlandse Organisatie voor Wetenschappelijk Onderzoek (NWO)'.
One of us, S.R., wishes to acknowledge fellowship support from the
Royal Netherlands Academy of Arts and Sciences.


\begin{thebibliography}{13}
\bibitem{footnote2} The region where the potential lowering due
  to the Coulomb potential of the dopant exceeds 0.1\,eV has a
  radius of about 1\,nm.
\end{thebibliography}
\end{document}